\documentclass[%
 reprint,
prb,
floatfix,
]{revtex4-2}

\usepackage{placeins}
\usepackage{natbib}
\usepackage{dcolumn}
\usepackage{bm}
\usepackage{graphicx}
\usepackage{wrapfig}
\usepackage{amsmath}
\usepackage{xspace}
\usepackage{capt-of}
\usepackage{float} 
\usepackage{epsfig}
\usepackage{rotating}
\usepackage{fancyhdr}
\usepackage[titletoc]{appendix}
\usepackage[english]{babel}
\usepackage{subcaption}
\usepackage[dvipsnames]{xcolor}

\usepackage[utf8]{inputenc}
\usepackage[T1]{fontenc}
\usepackage{mathptmx}
\usepackage{xcolor}

\begin{document}

\title{Spin-Phonon interaction in quasi 2D- $Cr_2Te_3$}

\author{Gurupada Ghorai}
\affiliation{School of Physical Sciences, National Institute of Science Education and Research (NISER) Bhubaneswar, An OCC of Homi Bhabha National Institute, Jatni-752050, Odisha, India.}
\author{Kalyan Ghosh}
\affiliation{School of Physical Sciences, National Institute of Science Education and Research (NISER) Bhubaneswar, An OCC of Homi Bhabha National Institute, Jatni-752050, Odisha, India.}
\author{Abhilash Patra}
\affiliation{School of Physical Sciences, National Institute of Science Education and Research (NISER) Bhubaneswar, An OCC of Homi Bhabha National Institute, Jatni-752050, Odisha, India.}
\author{Prasanjit Samal}
\affiliation{School of Physical Sciences, National Institute of Science Education and Research (NISER) Bhubaneswar, An OCC of Homi Bhabha National Institute, Jatni-752050, Odisha, India.}
\author{Kartik Senapati}
\affiliation{School of Physical Sciences, National Institute of Science Education and Research, An OCC of Homi Bhabha National Institute, Jatni, Odisha - 752050, India.}
\affiliation{Center for Interdisciplinary Sciences (CIS), NISER Bhubaneswar, HBNI, Jatni-752050, Odisha, India.}
\author{Pratap K. Sahoo}
\email{pratap.sahoo@niser.ac.in}
\affiliation{School of Physical Sciences, National Institute of Science Education and Research, An OCC of Homi Bhabha National Institute, Jatni, Odisha - 752050, India.}
\affiliation{Center for Interdisciplinary Sciences (CIS), NISER Bhubaneswar, HBNI, Jatni-752050, Odisha, India.}
\date{\today}

\begin{abstract}
Spin-phonon interaction plays an important role in 2D magnetic materials and motivates the development of next-generation spin- and charge-dependent microelectronic devices. Understanding the spin-phonon interaction by tuning the growth parameter of single crystal Cr$_2Te_3$, a robust quasi-2D room temperature magnetic material, is crucial for spintronic devices. The synthesis of single crystal 2D Cr$_2Te_3$ flakes on a Si substrate from co-deposited thin film by plasma annealing techniques is a significant achievement. The temperature dependence and polarization-resolved Raman spectroscopy with support of density functional theory classified lattice symmetry operations were used to identify the phonon modes to investigate the spin/electron-phonon interactions in Cr$_2Te_3$. The mean-field theory model in single crystal Cr$_2Te_3$ is employed to quantify the spin-phonon interaction and correlate with in-plane and out-of-plane magnetic behavior. The observation of a positive correlation between phonon mode frequency and spin-phonon interaction strength in single crystal Cr$_2Te_3$ can be a potential candidate for spintronic applications.

\begin{description}
\item[Keyword]
2D-Cr$_2$2Te$_3$, TEM, SQUID, Raman spectroscopy, DFT.
\end{description}
\end{abstract}

\maketitle

\section{INTRODUCTION}
The thermal transport and spin properties of the 2D magnetic materials are crucial for spintronic devices\cite{1}. Although spin-phonon, electron-phonon, and phonon-phonon interactions are offered for study for high-performance spin-based thermal management devices\cite{Ahn2020,doi.org/10.1016/j.ssc.2018.08.008}. This recommends establishing the influence of spin-phonon interaction on the physical phenomena of the 2D magnet. The spin-phonon solid interaction in the 2D $CrSiTe_3$ was proposed from the temperature-dependent I-R spectroscopy and super-exchange mechanism influenced in the Si-Te stretching and Te displacement in temperature-dependent frequency study\cite{Milos, CST1}. The unconventional spin-phonon interaction vis Dzyaloshinskii–Moriya interaction(DMI) in $Y_2Ir_2O_7$ crystal by temperature-dependent infrared spectroscopy was also reported\cite{https://doi.org/10.1038/s41535-019-0157-0}. The spin-phonon interaction in the bulk crystal of yttrium iron garnet (YIG) with enhanced interaction strength at higher phonon frequency was investigated \cite{PhysRevB.104.L020401}. However, spin-phonon coupling is a primary need to tackle how spin affects phonon thermal transport. To address these coupling issues, we are interested in exploring the spin-phonon interaction of $Cr_2Te_3$. The 2D magnetic order in chromium trihalide materials was first discovered\cite{Seyler}. The 2D magnetic materials could have been lead to one of the spintronics materials by utilizing charge and spin for applications ranging from molecular quantum devices and sensing to ultrathin high-density data storage devices\cite{9,10}. In a similar line, the intrinsic 2D layer-dependent ferromagnetism properties in CrGe$_2$ were also investigated \cite{nature}. The structural properties using scanning tunneling microscopy, bandgap, and electronic properties were calculated using DFT calculation of $CrGeTe_3$ materials in ref\cite{Zhenqi}. Although three. Theoretically,  it has been suggested that layered materials like $CrXTe_3(X=Si, Ge, Sn)$ show 2D ferromagnetic or antiferromagnetic behavior dependent on their exchange or superexchange interaction strength\cite{Huang2018,C5TC03463A}. It has been proposed that $Cr_2Te_3$ is a tunable curie temperature up to room temperature and suitable for magnetic memory device applications\cite{Wen2020}. So, for magnetic-based applications of $Cr_2Te_3$, there should be clear evidence of the magnetic properties of such material, which are influenced by defect spins in the lattice. It has been demonstrated that the 2D $Cr_2Te_3$ materials show ferrimagnet or ferromagnet properties, which are dependent on the strength of the nearest neighbor exchange interaction\cite{doi:10.1063/1.2713699}. The (001) oriented crystal grown at a temperature of 400$^\circ$C at a rate of ~1.7 nm/min on Al$_2O_3$(0001) using molecular beam epitaxial(MBE) reveals the local electronic and magnetic structure of Cr ions, which was demonstrated the edge of energy of Cr L$_{2,3}$ using X-ray magnetic circular dichroism (XMCD) and soft x-ray absorption spectroscopy\cite{Burn2019}. However, the magnetic structure of $Cr_2Te_3$ is not yet fully understood, because of the complicated coupling of the spins in three distinct Cr sites. Curie temperature and magnetic properties vary with various defect sites in the 2D magnetic materials.

 To understand the spin-phonon interaction, we have synthesized a single crystal $Cr_2Te_3$, a quasi 2D magnetic material by co-deposited Cr and Te thin film followed by thermal annealing. The $A_g$ and three $E_g$ phonon modes were analyzed from the temperature-dependent Raman study. Different vibrational modes and frequencies were obtained using DFT calculations and compared with the experimental results. The anharmonicity of Raman modes can be well-fitted with Balkanski's model above the critical temperature. The ferromagnetic transition with temperature matched well with the Raman anharmonicity, which can be explained in terms of spin-phonon interaction. The results can be understood in terms of magnon-phonon coupling.

\section{Experimental method}
The commercially available high-purity (99.999$\%$, Sigma Aldrich) metal pieces of Chromium(Cr) and Tellurium(Te) are co-evaporated using the e-beam evaporation techniques and thermal evaporation technique in a base vacuum of 2$\times10^{-7}$ mbar to produce thin films of CrxTey on the clean Si substrate. The deposition rate of Cr and Te were 0.1 \AA/s and 0.4\AA/s, respectively, and the thickness of 40 nm was measured by a quartz crystal thickness monitor attached to the deposition chamber. The Co-deposited samples were annealed at 500$^\circ$C inside of quartz tube of the plasma-enhanced chemical vapor deposition (PECVD) system at 120 W plasma power with the ambient of forming gas for two hours. The surface micrographs were collected using a Field emission of scanning electron microscopy (FESEM) system (SIGMA ZEISS), and the elemental mapping and percentage of composition were estimated using Apex energy dispersive spectroscopy (EDS) attached to the FESEM. The XRD pattern of the $Cr_2Te_3$ crystalline phases was obtained using a Rigaku diffractometer with $CuK_\alpha$ radiation ($\lambda=1.54056$ \AA) at room temperature with a grazing incidence X-ray diffraction (GIXRD) approach at a grazing angle of 1.0 degree with a step size of 0.02deg. The local crystalline structural properties of the 2D-$Cr_2Te_3$ were verified using a high-resolution transmission electron microscope (HRTEM) of a 200 KV energetic electron source. The high-resolution confocal Raman scattering with a spatial resolution of 2-4 microns was used to understand the vibrational modes of $Cr_2Te_3$. The temperature-dependent Raman spectrum of $Cr_2Te_3$ flake was measured using a Linkam cooling stage. The specimens were mounted on a sample holder fixed to a cold head in a vacuum with a constant liquid nitrogen flow to cool the chamber down to 80K. A half-wave plate in an incident light path was fixed to study the polarization-dependent Raman spectra. We also calculated the Raman peaks position, and modes of vibration were obtained theoretically using density functional theory (DFT). The temperature dependence of the Raman spectrum of $Cr_2Te_3$ flakes was measured using a Horiba Raman microscope equipped with a Linkam cooling stage. The specimens were mounted on a non-background sample holder fixed to a cold head in a vacuum, and the liquid nitrogen was used to cool the chamber up to 80K. The superconducting quantum interference device (SQUID) system of Quantum Design was used to investigate the magnetic properties of $Cr_2Te_3$ in the temperature range of 5-250K.

\section{Results and discussions}

\begin{figure*}[!htbp]
\includegraphics[width=6in]{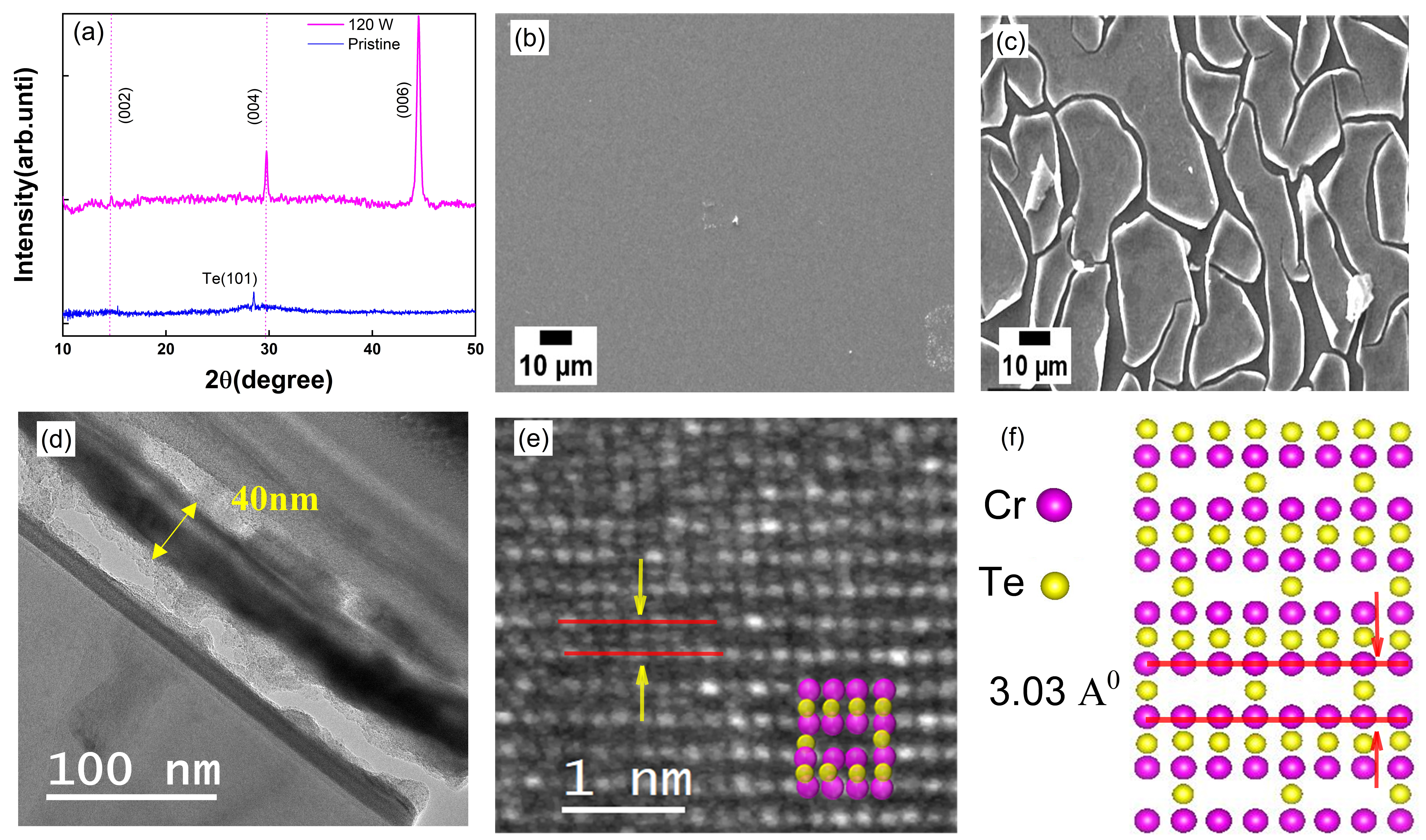}
\caption{\label{fig: wide1} (a) GIXRD data of the pristine, applied  120 W plasma power samples. (b) Schematic illustration of crystal structure of $Cr_2Te_3$. (c)  FESEM image of co-deposited films,  (d) Surface morphology image of $Cr_2Te_3$ sample after annealed $500^0C$ with the presence of 120 W plasma power.}
\label{XRD1}
\end{figure*}

The GIXRD measurement at a grazing angle of 1$^\circ$ was used to obtain the crystalline properties and the orientation of the $Cr_2Te_3$ growth. The GIXRD spectra of the pristine and 120W plasma, 500$^\circ$C annealed samples are shown in figure \ref{XRD1}(a). The pristine sample shows a weak XRD peak at 27.56$^\circ$, corresponding to the lattice plane of Te (101) (JCPDS \# 36-1452)\cite{https://doi.org/10.1016/j.jallcom.2015.02.007,10.1039/C9RA08692G}. The 120W plasma annealed sample shows three XRD peak positions at 2$\theta$ values of 14.54$^\circ$, 29.68$^\circ$, and 44.52$^\circ$ corresponding to the lattice plane of (002), (004), and (006), respectively for the crystalline plane of $Cr_2Te_3$. The above lattice plans describe the Trigonal family of $Cr_2Te_3$, with the space group of 163 $(P\bar{3}1c)$, with lattice parameters of a=6.78 \AA, b=6.78 \AA, c=12.06 \AA, $\alpha$=90, $\beta$=90 and $\gamma$=120 (PDF\#71-2245), which matches well with an earlier report\cite{doi:10.1021/acs.nanolett.9b05128, Burn2019}. Figure \ref{XRD1}(b) and (c) shows the FESEM surface morphology of the pristine and 120 W annealed sample. The pristine sample shows a very smooth surface. It has been observed that after annealing the pristine at 500$^\circ$C in 120W plasma for 2 hours, the CrTe thin films grew as individual flakes with clear grain boundaries, as shown in figure \ref{XRD1}(c), which is a 2D $Cr_2Te_3$.  To further clarify the crystalline growth of quasi 2D $Cr_2Te_3$, we have prepared a lamella for TEM study from a single flake. The lamella was prepared using a focused ion beam (FIB) system using 30 keV Ga beam followed by in situ polishing by low energy Ar ion. The lamella provides the cross-sectional view of 40 nm thin flake as shown in figure \ref{XRD1}(d), which is basically the c-axis of the 2D-$Cr_2Te_3$ crystalline plane. The high-resolution TEM with lattice arrangements and corresponding selected area electron diffraction (SAED) patterns of 2D-Cr$_2Te_3$ is shown in figure \ref{XRD1}(e). The yellow parallel line shows the (001) plane of the Cr$_2Te_3$ with interplanar separation d$_{001}$ of 3.03 \AA. The schematic image along the (010) zone axis and the lattice spacing of d$_{001}$ is shown in figure \ref{XRD1}(f). The XRD and HRTEM images strongly suggest the crystalline growth of quasi-2D $Cr_2Te_3$ along the (001) plane.

\subsection{Raman spectra of Cr$_2$Te$_3$ from density functional theory}
\begin{figure*}[!htbp]
\includegraphics[width=6in]{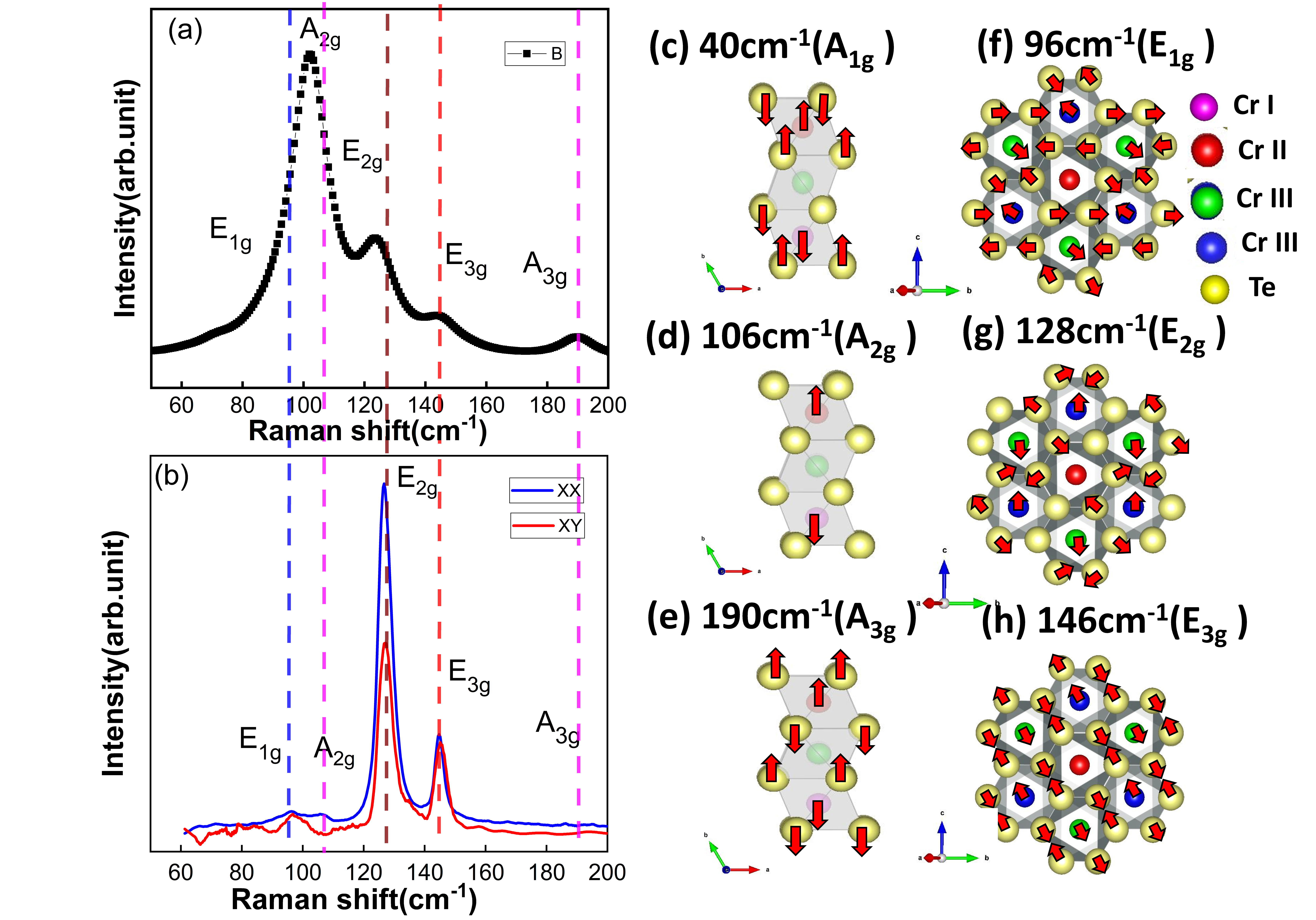}
\caption{\label{fig:wide2} (a) Polarisation dependent Raman at the temperature 150K where blue color code for unpolarised(XX) condition and red color curve for polarised(XY) Raman spectrum. (b) The theoretical DFT calculation of the Raman spectrum of $Cr_2Te_3$ single crystal. The atomic displacements of the Raman active modes (the color code of atoms is the same as Fig 2(a)). The out-of-plane $A_g$ vibrational modes are depicted on the left panel (top view and side view), while the in-plane$E_g$ vibrational modes are on the right (top view only). }
\label{atomic vibration}
\end{figure*}

Raman spectroscopy is a versatile tool to understand the electron-phonon or spin-phonon interaction in 2D materials. Depending on the crystal orientation the vibrational modes can be suppressed or enhanced which may affect the spin-phonon interaction. The observation of Raman modes in the case of Cr$_2Te_3$ strongly depends on the crystalline quality and synthesis process [xx-xx]. In order to identify different Raman modes of Cr$_2Te_3$ , we have used the density functional theory\cite{hohenberg1964inhomogeneous, kohn1965self} using the Vienna Ab-initio Simulation Package (VASP).\cite{vasp-1,vasp-2,vasp-3,vasp-4}. We have used the recommended Perdew-Burke-Ernzerhof\cite{perdew1996generalized} (PBE) PAW pseudo potentials in all the calculations along with the plane wave basis cutoff of 500 eV and set the value of EDIFF = $1.0 \times 10^{-8}$ eV to represent an energy convergence criterion. For the Brillouin zone integration, a $\Gamma$-centered ($12 \times 12 \times 6$) k-point sampling is used. The Hubbard correction method within semilocal DFT\cite{Cococcioni2005linear}, i.e., GGA(PBE)+U is used for both structure optimization and studying the Raman spectra. The value of U=3.0 eV\cite{lee2021modulating} for Cr atoms is set to take into account the orbital localization. The unit cell of Cr$_2$Te$_3$ with P-31c space group is optimized with the above given set up and obtained the lattice constants as a=b=6.99 \AA, and c=12.65 \AA, respectively. 
To get the Raman active modes, the dielectric matrix is calculated for each mode using the density functional perturbation theory(DFPT) as implemented in the VASP package. The intensity of each mode is calculated following the Ref. \cite{porezag1996infrared} and using the raman-sc\cite{vasp$_$raman$_$py} python package. The Raman spectra of Cr$_2$Te$_3$ is shown in figure \ref{atomic vibration}(a). The Lorentzian broadening width of 5 cm$^{-1}$ is used to match with the experimental spectra. Six dominating peaks corresponding to more intense peaks are found at 40, 73, 102, 123, 144, and 190 cm$^{-1}$. For comparison, the experimental Raman peaks are shown in figure \ref{atomic vibration}(b). The peaks at 102, 123, and 144 cm$^{-1}$ are closer to experimental peaks at 106, 128, and 146 cm$^{-1}$. One can observe that two modes at 40 and 190 cm$^{-1}$ are not observed in the experimental Raman spectra. The Raman modes observed from the experiment and DFT calculation are tabulated in Table \ref{sec:level21}.

It should be noted that Cr$_2$Te$_3$ belongs to the point group of D$_{3d}$(-3m). Based on the point groups of the character table, the deduced Raman active modes for Cr$_2$Te$_3$ are 4A$_{1g}$+9E$_g$. The Raman-active mode intensity is related to the Raman tensor and the direction of incident and scattered light as: $I_m^R=\frac{d\sigma}{d\Omega} \propto \sum_m^{ } |\Vec{e}_s \cdot Rt_m \cdot \Vec{e}_i|^2$. Where Rt$_m$ is the Raman tensor of m$^{th}$ active mode and $\Vec{e}_s$ ( $\Vec{e}_i$) is polarization vectors of scattered (incoming) light\cite{zhang2016review,niranjan2021significance}. The Raman tensors correspond to active Raman modes A$_{1g}$, and E$_{g}$ are as follows,  
\begin{center}
A$_{1g}(D{_{3d}}(-3m)) : $
$\begin{pmatrix}
a&0&0\\
0&a&0\\
0&0&b
\end{pmatrix}$
E$_{g}(D{_{3d}}(-3m)):
\begin{pmatrix}
c&0&0\\
0&-c&d\\
0&d&0
\end{pmatrix},
\begin{pmatrix}
0&-c&-d\\
-c&0&0\\
-d&0&0
\end{pmatrix}.$
\end{center}
We use the notation XX and YX to represent unpolarized and polarized scattering configurations, respectively. In the case of unpolarized configuration XX, both the incoming and scattering polarization vectors are $\Vec{e_i}$=(1, 0, 0)=$\Vec{e_s}$ and for YX, $\Vec{e_i}$=(1, 0, 0), $\Vec{e_s}=(0, 1, 0)$. So, under unpolarized configuration XX, I(A$_{1g}$) = a$^2$, and I(E$_g$) = c$^2$. Under polarized condition YX, I(A$_{1g}$) = 0, and I(E$_g$) = c$^2$. So, in cross polarization condition, the intensity of A$_{1g}$ mode vanishes, and E$_g$ peaks remain unchanged. 

\begin{table}[!htbp]
\caption{\label{sec:level21}The experimental and theoretical frequencies, and vibrational modes of strain-free R3 Cr$_2Te_3$. The ‘$||$’ and ‘$\perp$’ represent the parallel and cross polarization conditions, respectively. The ‘$-$’ represents peaks that were not clearly observed.}
\begin{ruledtabular}
\begin{tabular}{ccccccc}
mode&$A_g$&$E_g$&$A_g$&$E_g$&$E_g$&$A_g$\\\hline
EXPT‘$||$’&\mbox{-}&\mbox{96}&\mbox{106}&\mbox{128}&\mbox{146}&\mbox{-}\\
EXPT‘$\perp$’&\mbox{-}&\mbox{96}&\mbox{-}&\mbox{128}&\mbox{146}&\mbox{-}\\
Theory&\mbox{40}&\mbox{73}&\mbox{102}&\mbox{123}&\mbox{144}&\mbox{190}\\
\end{tabular}
\end{ruledtabular}
\end{table}

The Raman active modes for $Cr_2Te_3$ belong to the trigonal lattice structure with 163 space group\cite{szhou}. For this space group, two types of active Raman vibration modes are observed, i.e  non-degenerate $E_g$ and  degenerate $A_g$\cite{szhou}. So the $A_g$ mode is the out-plane lattice vibration and $E_g$ is in-plane atomic vibration of the lattice.

We have studied the polarization Raman spectra at low temperatures (80K) to best match the theoretical Raman modes. Compared to theoretical parameters where unpolarized configuration XX, I(A$_{1g}$) = a$^2$, and I(E$_g$) = c$^2$, the experimental spectra shows four Raman peaks at $96cm^{-1}$, $106cm^{-1}$, $128cm^{-1}$ and $146cm^{-1}$. For the theoretical polarized condition YX, I(A$_{1g}$) = 0, and I(E$_g$) = c$^2$, the experimental spectra shows three Raman peaks at $96cm^{-1}$, $128cm^{-1}$ and $146cm^{-1}$. The above observation clearly illustrates the condition and provides the information that the $A_g$ is suppressed and other peaks correspond to $E_g$ Raman modes. This procedure quantifies the modes of vibration in $Cr_2Te_3$ quasi 2D materials. The schematic configuration of the atomic vibration in the lattice atoms of $Cr_2Te_3$ is shown in figure \ref{atomic vibration}(c, d, e, f, g, h). The direction of vibration is shown in red arrows and the Cr and Te atoms are shown in magenta and yellow, respectively. Figure \ref{atomic vibration}(c, d, e) shows the three configurations as A$_{1g}$, A$_{2g}$, A$_{3g}$ for out-of-plane vibration in the c-axis, in which the light interact ‘$||$’ to the lattice plane. Figure \ref{atomic vibration}(f, g, h) represents three configurations as E$_{1g}$, E$_{2g}$, E$_{3g}$ for in-plane vibration in the XY plane, in which polarized light interacts $\perp$ to the lattice plane.

\subsection{Temperature dependent Raman study:}
The interaction of phonon with spin and magnon can be understood from the temperature dependant Raman spectroscopy. The interpretation of Raman spectra of soft modes in a ferromagnetic material can be understood from its over-damped line shape and the temperature dependence near the transition temperature ($T_N$).It has been observed that Raman spectroscopy a unique tool to probe spin-phonon coupling(s) which is strongly influenced by elementary excitations in multiferroic materials \cite{PhysRevB.104.L020401}. For a better understanding of the phonon modes in Cr$_2$Te$_3$, systematic Raman spectra were collected in the temperature range of 80-300 K and analyzed their evolution using Lorentzian multifunction. The temperature dependant Raman spectra of Cr$_2$Te$_3$, sample synthesized at 120 W is shown in Fig. 3(a). Two distinct peaks at 125 cm$^{-1}$ and 144 cm$^{-1}$ corresponding to $E_{2g}$ and $E_{3g}$, respectively,  are observed with peak intensities varying with temperature. A 2D contour plot shown in Fig 3(b) indicates the distribution of Raman peak intensities within the frequency range of 80-165 cm$^{-1}$. One can clearly perceive that both $E_{g}$ mode intensities are maximum within the temperature range of 100-165 K and then reduced at the higher temperature. The evolution of parameters like peak positions, full width at half maximum (FWHM), and integrated intensity as a function of temperature are evaluated to understand the strength of the spin-phonon interaction and the phonon dynamics of the system. 

\begin{figure}[!htpb]
\includegraphics[width=3.2in]{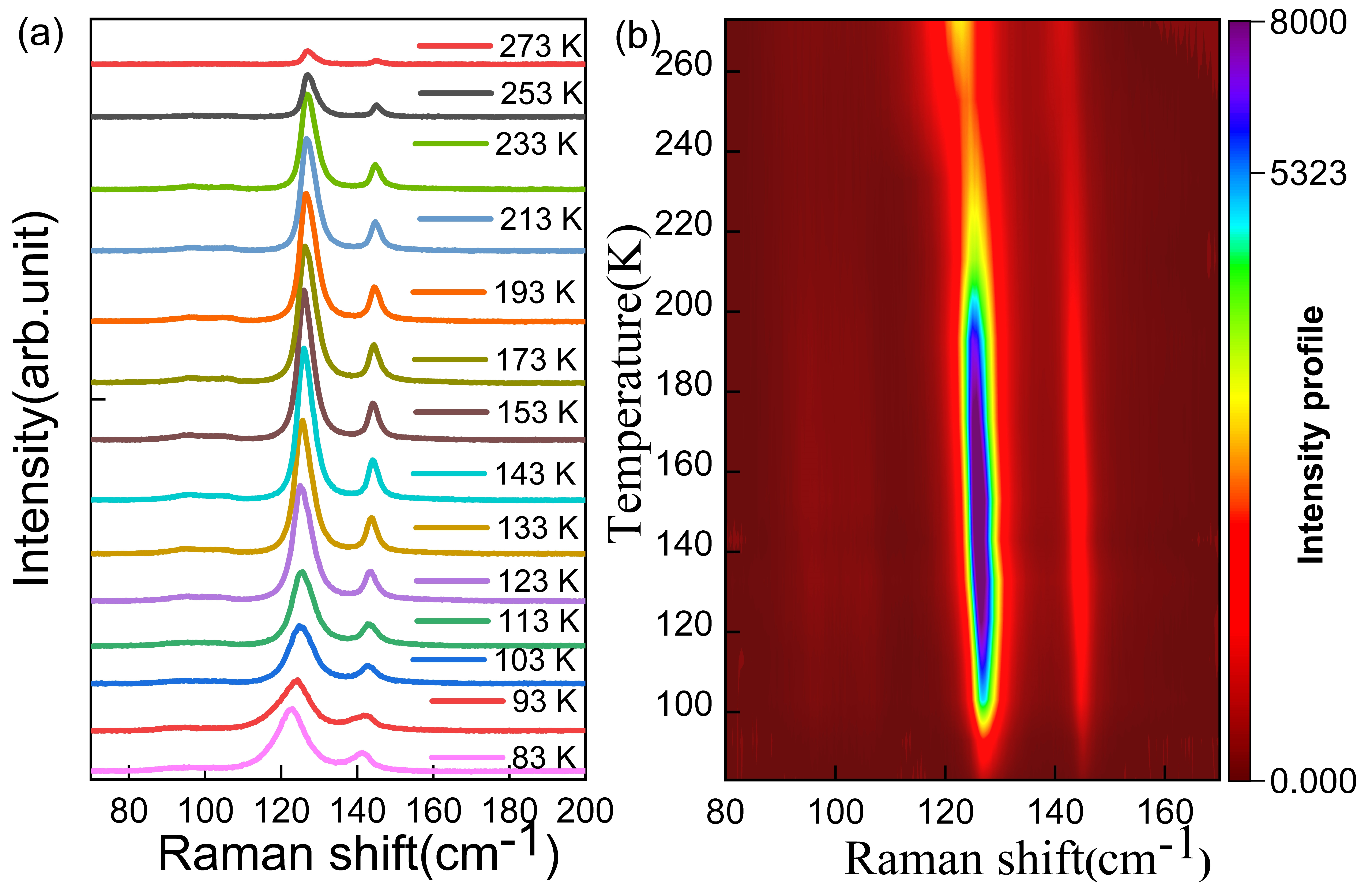}
\caption{(a) Temperature dependent Raman spectrum of 120 W plasma power samples in the temperature range of 82K to 273K. (b) 2D contour, in the range of
80 to 270 cm$^{-1}$. \label{fig:wide} }
\label{atomic vibration1}
\end{figure}

\begin{figure*}[!htbp]
\includegraphics[width=5in]{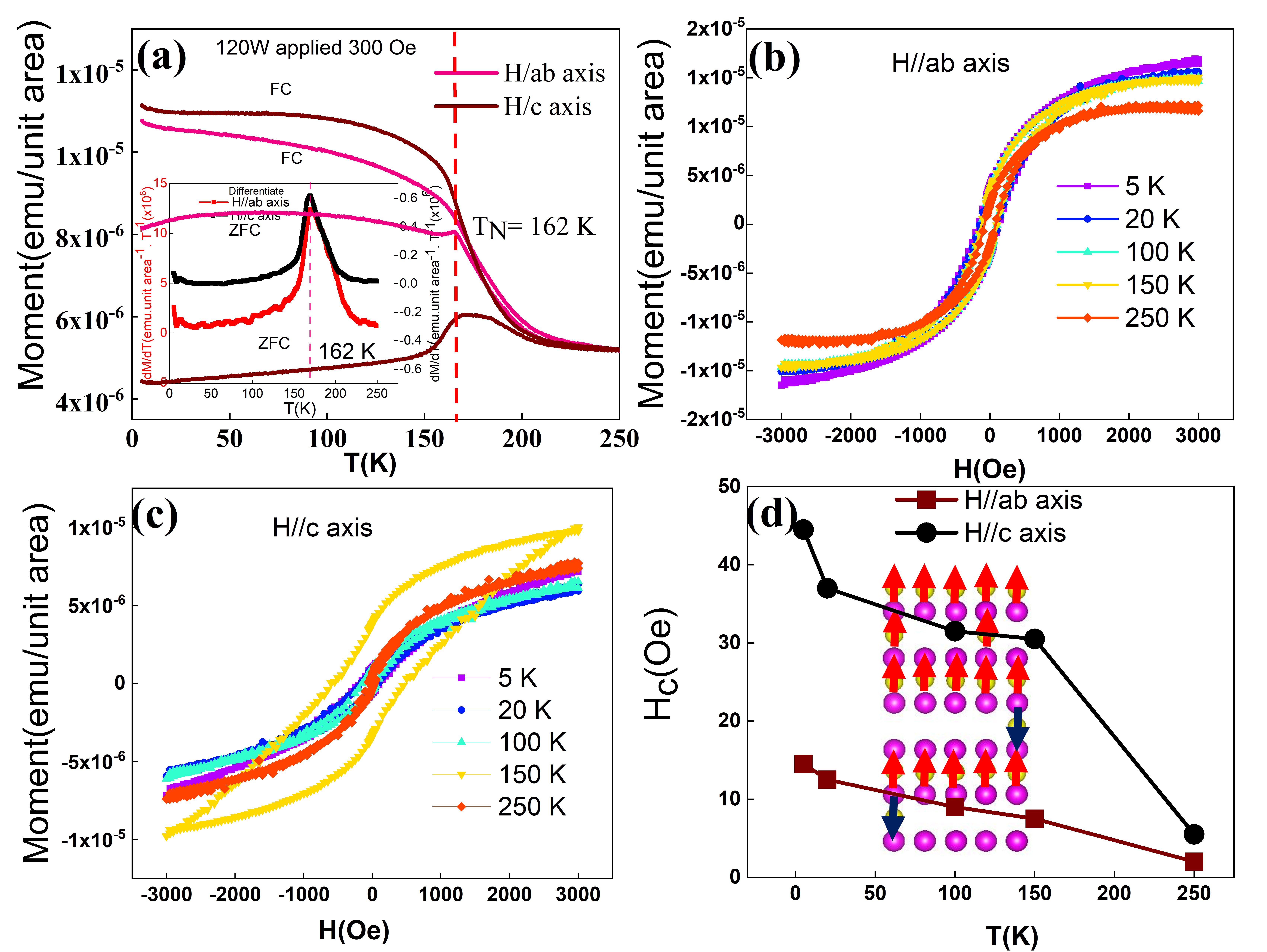}
\caption{Magnetisation measurement (a) The MH loop measurement parallel the ab plane configuration of 120 W plasma sample. (b) The MH loop parallel to c axis of the 120 W plasma sample. (c) The MT measurement with applied 300 Oe field parallel and perpendicular to the ab plane of 120 W plasma used sample.(d) are the Exchange bias field of the above four configurations.}
\label{ZFC FC}
\end{figure*}

\subsection{Magnetization study:}

In the magnetically ordered system, both Raman line shift and magnetic moments deviate at similar temperatures due to the strong spin-phonon interaction. In order to correlate that, the magnetic properties of the material, hysteresis loop, zero field cooled (ZFC), and field cooled (FC), measurements were performed with both perpendicular and parallel applied external magnetic fields with respect to the sample plane. In Figure \ref{ZFC FC}(a), the magnetic moments (ZFC and FC) are plotted as a function of temperature for the 120 W plasma power sample. The brown curve represents the case of a perpendicular applied field of 300 Oe to the ab plane, while the pink curve represents the case of an applied field parallel to the ab plane. The inset of Figure \ref{ZFC FC} shows the first order derivative of M-T, and the T$_C$ is determined to be $\approx162K$. However, the ZFC curve of the parallel external magnetic field for the 120 W plasma power samples shows an increase in the magnetic moment at the transition temperature. This suggests the presence of a well-defined paramagnetic to AFM phase transition. In the case of $Cr_2Te_3$, the magnetic behavior is complex due to the coexistence of strong FM and weak AFM interactions within the system. The system exhibits spin frustration at lower temperatures, leading to the emergence of a re-entrant spin glass phase. Previous studies by Roy et al. and Hashimoto et al. have reported the presence of both FM and AFM interactions in $Cr_2Te_3$, with the stronger FM interactions dominating above the T$_C$ (165K)\cite{Roy2015}. The existence of a more complex magnetic order below the T$_C$ has also been predicted. The magnetic field-dependant magnetization M-H measurement of the 120W plasma power sample was performed to confirm the FM and perpendicular anisotropy.  The M-H loops are shown in Figure \ref{ZFC FC}(b) and (c). Figure \ref{ZFC FC}(d) shows the exchange bias field from the M-H loop as a function of temperature.

It has been observed that the exchange field decreases with increasing temperature, indicating the presence of FM and AFM coupling in the sample. The observation of exchange bias in $Cr_2Te_3$ coupled with the ferromagnet CdTe suggests the presence of an AFM component in $Cr_2Te_3$, likely arising from c-axis Cr atom defects in the lattice\cite{doi:10.1063/1.3677883}.  The existence of both ferromagnet or ferrimagnet properties in $Cr_2Te_3$ dependent on the nearest neighbor spin coupling energy of the Cr sites \cite{doi:10.1063/1.2713699}. It has been reported that the ferromagnetic order enhanced perpendicular magnetic anisotropy. Depending on the spin orientations by the applied magnetic field, the net magnetization in both directions is different. So, the magnetization property will depend on the direction of the applied magnetic field to the c-axis of the $Cr_2Te_3$.  The unit cell of $Cr_2Te_3$ contains alternating layers of Cr and Te atoms and has Cr$_I$ vacancies in every second metal layer. The Cr$_I$ atoms in the partially occupied layers are sandwiched between the Cr$_{III}$ layers and have no close-by neighbors in the a-b plane, whereas the Cr$_{II}$ and Cr$_{III}$ atoms make up the fully occupied metal layers. The crystal growth resembles a hexagonal lattice in the (001) plane, indicating high crystalline quality as observed from XRD measurements.

\subsection{Spin-phonon interaction in $Cr_2Te_3$}

\begin{figure}[!ht]
\centering
\includegraphics[width=3.3in]{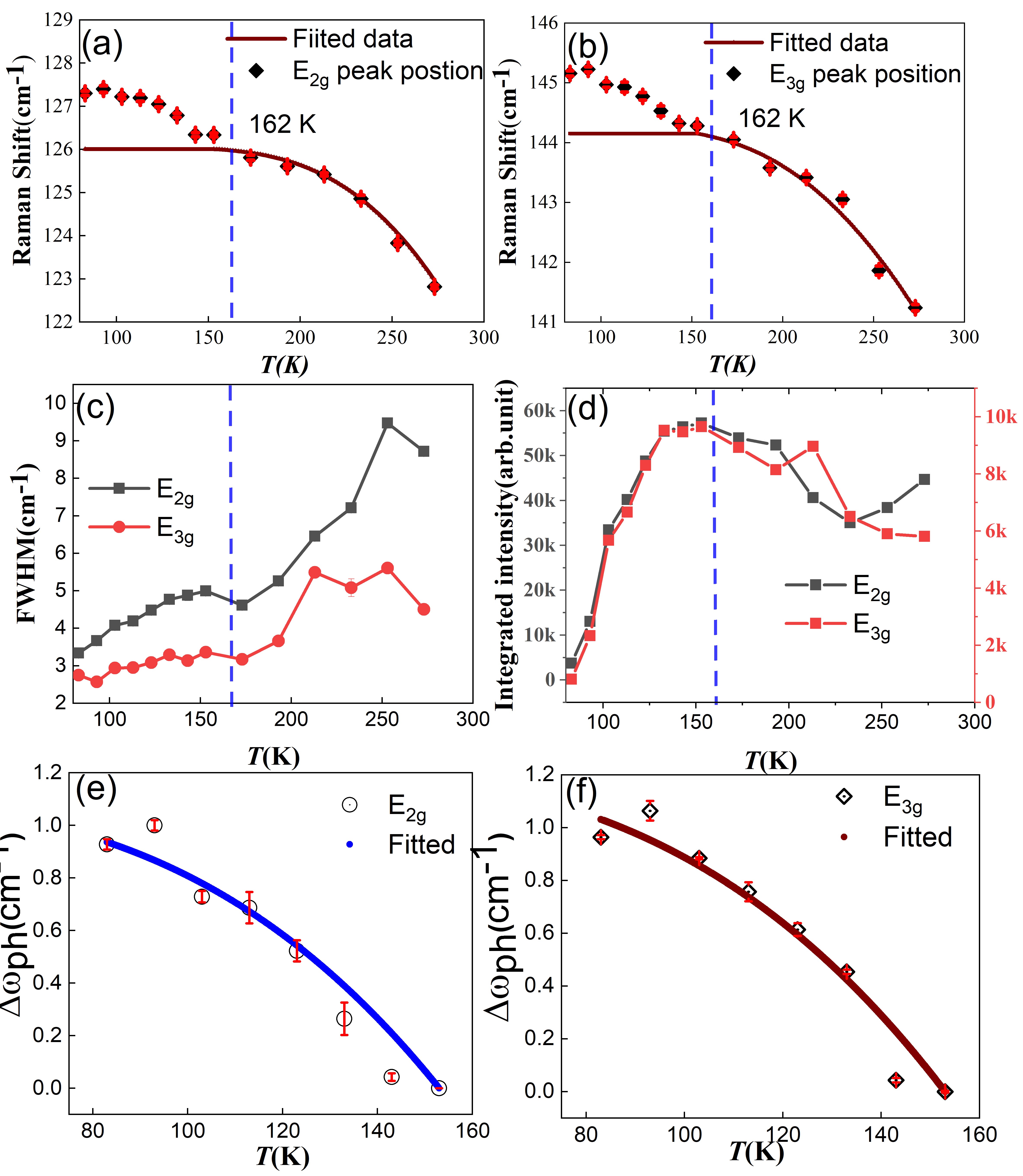}
\caption{Temperature dependent Raman spectrum, (a) The Raman peak position of E$_{2g}$ (125 cm$^{-1}$) and (b) E$_{3g}$ (145 cm$^{-1}$) mode of vibration with temperature and fitted equation(1) for showing the anharmonicity. (c) FWHM with the function of temperature of two most prominent peaks of E$_{2g}$ and E$_{3g}$ modes, respectively. (d) Integrated intensity of Raman peaks as a function of temperature of two most prominent peaks of E$_{2g}$ and E$_{3g}$ modes. The extra Raman peak shift below the transition temperature of both peaks (e) E$_{2g}$ and (f) E$_{3g}$ and fitting function equation \ref{Klemens}. }
\label{tem_raman}
\end{figure}

The Raman shift of the most prominent peaks, $E_{2g}$(125 cm$^{-1}$) and $E_{3g}$(144 cm$^{-1}$), as a function of temperature are shown in Figure \ref{tem_raman}(a) and (b). The Balkanski model fits the data well, indicating the presence of strong electron-phonon interactions. The FWHM and integrated intensity of the $E_{2g}$(125 cm$^{-1}$) and $E_{3g}$(144 cm$^{-1}$) peaks were studied as a function of temperature, as depicted in Figure \ref{tem_raman}(c) and (d). Anomalous behavior near the transition temperature of 166 K indicates that all Raman peaks correspond to the quasi-2D $Cr_2Te_3$ film. Magnetization measurement of $Cr_2Te_3$ confirms the presence of a strong ferromagnetic component with a weak antiferromagnetic component. The temperature-dependent behavior of the magnetic moment suggests the induction of a magnetic moment below the transition temperature, owing to the influence of nearest neighboring spin-spin interactions. The excellent agreement between the temperature-dependent Raman spectroscopy and magnetization measurements, and theoretical calculations provides a comprehensive understanding of the spin-phonon interaction in $Cr_2Te_3$, which is rucial for potential applications in spintronics and contributes to the understanding of the material's behavior and properties.

In the absence of spin order above the transition point, the temperature-induced phonon position shift follows a three-phonon scattering model that takes into account the anharmonicity contributions leading to softening or hardening of some Raman modes. The occupation probabilities of the phonons are responsible for the temperature dependence Raman shift and linewidth broadening. In the case of the phonon decay process, the optical phonon at $\gamma$ point with energy $\hbar \omega_{c2}(0)$ decays into two acoustic phonons from the same branch, keeping both energy and momentum conserved. So the decay phonon has $\hbar\omega_{c2}(0)$/2 energy with equal and opposite momentum. This process can be expressed as the temperature dependence Raman shift for the optical phonon frequency $\omega_{c2}(T)$  as follows (Klemens’s model): 
\begin{equation} 
\omega_{c2}(T)= \omega_{0}- C(1+\frac{2}{\exp\frac{\hbar\omega}{k_BT}-1})
\end{equation}
Here, $\omega_0$ represents the phonon frequency at 0 K, $\hbar$ is the reduced Planck constant, $k_B$ is the Boltzmann constant, T is the temperature, and C is the three phonon interaction constant. The first term represents the initial phonon frequency, while the second term accounts for anharmonic contributions due to phonon decay. This can be understood by fitting parameters such as C and $\omega_0$. 
It has been observed that the temperature dependence Raman shift fitted well with Klemens's model up to 162K. The anomaly below 162K does not follow the three-phonon decay process. According to Klemens's model, the decayed acoustic phonon should have a constant velocity below the critical temperature depending on the thermal conductivity of the materials. However, the upturn behavior indicates the increase of the phonon velocity. By further decreasing the temperature below 162 K the Raman shift deviats from the expected anharmonic tendency indicating an additional scattering mechanism involved. Such a mechanism can be attributed to the coupling of the lattice to spin fluctuations. e.g., spin-phonon interaction. Revisiting the M-T data of $Cr_2Te_3$, it has been observed that the paramagnetic to FM transition occurs at 162K, indicating the role of spin below T$_C$ of 162K. Fennie et al. suggested that for an FM material spin coupled with optical phonon below transition temperature \cite{PhysRevLett.96.205505}. The optical phonon frequency $\omega{'}$ can be considered as the contribution of the anharmonic behavior and an extra term due to the deviation due to spin-phonon interaction ($\lambda_{sp}$), which can be written as :
\begin{equation} 
\omega{'}= \omega_{c2}(T)+ \Delta{\omega_{sp}}
\end{equation}

Where $\omega_{c2}(T)$ is the Raman frequency shift as mentioned in the Klemens model and $\Delta\omega_{sp}$ is the deviation due to spin-phonon interaction from the Klemens model fitting. Finally, the spin-phonon coupling parameter $\lambda_{sp}$ for $Cr_2Te_3$ can be calculated using this method. This effect is similar to that observed in other magnetic materials, and it is associated with phonon renormalization induced by magnetic ordering due to the coupling between magnetic ordering and crystal lattice. Considering the N-N Heisenberg spin system, the magnetic energy of the FM system can be correlated with the frequency shift of lattice vibrations by introducing the proportional constant as spin-phonon coupling parameter $\lambda_{sp}$, which depends on the derivatives of the exchange constants with respect to the coordinates of the magnetic ions. In the absence of magnetostriction effects and electronic states renormalization, the contribution of spin-phonon coupling $(\lambda_{sp}$ of the $k^{th}$ position is approximately given by\cite{PhysRevB.104.L020401, PhysRevB.105.075145}:

\begin{equation}
(\Delta{\omega_k})_{sp}=-\sum_{i,j}\lambda_{sp}<S_i.S_j>
\end{equation}

where $<S_i.S_j>$  is the spin correlation function. This contribution, considering only first-neighbor interactions and a molecular field approximation, can be considered in the case of the lattice vibration, as

\begin{equation}
(\Delta{\omega_{p}})=-\lambda_{sp}S^{2}\phi(T)
\label{Klemens}
\end{equation}

where $\Delta\omega_{p}$ indicates the deviation of Raman frequency, $\lambda_{sp}$ is the coefficients of the spin-phonon coupling, $<S_i.S_{i+1}>$ denotes the average for adjacent spins (where S can be taken as 3/2 for Cr spins); and $\phi$(T) is the order parameter given by $\phi(T)=1-(T/T_\theta)^\gamma$ where $T_\theta$ is the transition temperature. The temperature dependence function of $\phi(T)$ is the normalized order parameter, which is proportional to the square of the average magnetization M(T).

\begin{figure}[!ht]
\centering
\includegraphics[width=3.3in]{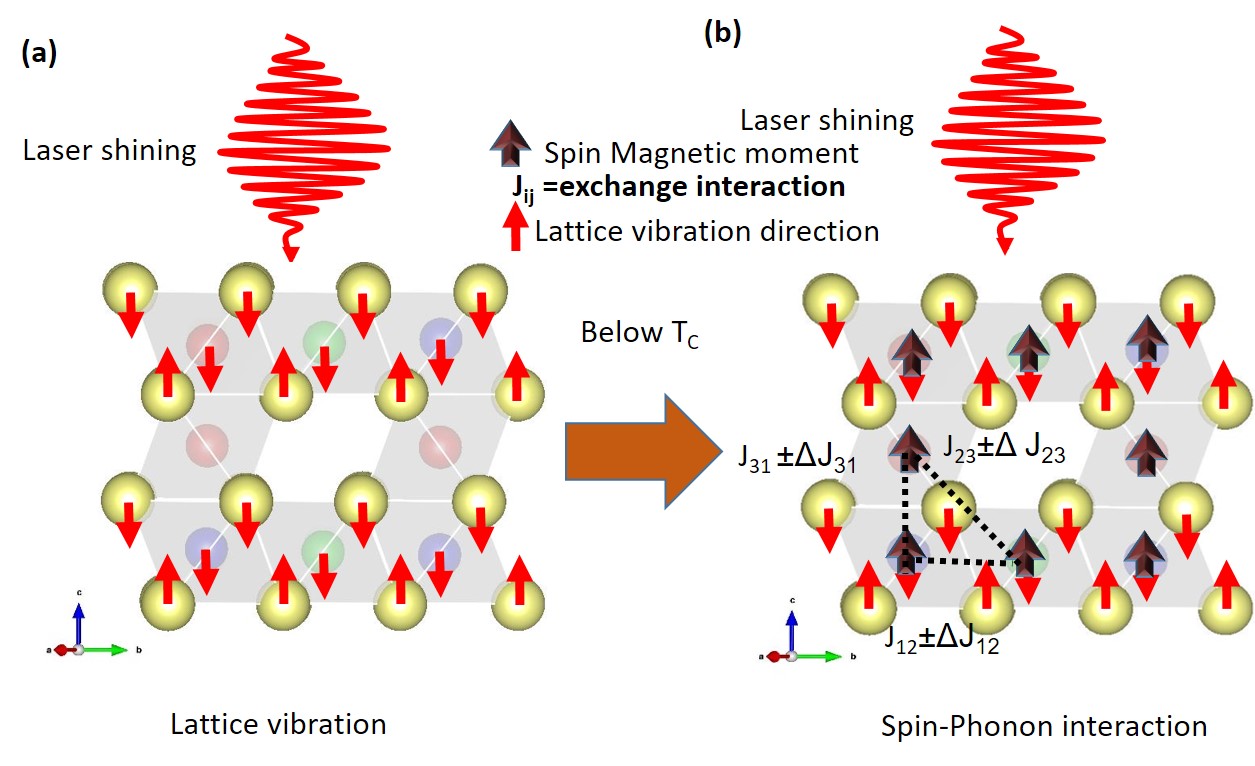}
\caption{(a) Above the T$_C$ the lattice vibration are observed due to Laser shining. (b) Below the transition temperature lattice vibration perturbed the exchange interaction.}
\label{model}
\end{figure}

The quasi-2D ferromagnetism in $Cr_2Te_3$ is intriguing due to its high spin-phonon interaction, especially above its paramagnetic region. Klemens's model well fits the temperature dependence Raman frequency till T$_C$, $\approx $162K and anomalies below that, as shown in Figure \ref{tem_raman}(a) and (b), indicating anharmonicity and spin-phonon interactions. Figure \ref{tem_raman}(e) and (f) fitted with the equations 3 and 4 in the temperature range of 80 – 160 K. The spin-phonon interaction parameter ($\lambda_{sp}$) is calculated from the fitting and found to be 0.47$\pm$0.03 cm$^{-1}$. This value confirms the presence of a strong spin-phonon interaction, which is responsible for the large phonon dispersion. Below the transition temperature, the order parameter is well-fitted, and the $\gamma$ value is determined to be 2.61$\pm$0.04. The value of $\lambda_{sp}$ is comparable with the literature and indicates the colinear spin arrangement describes the magnetostructural transition below T$_C$. The temperature-dependent Raman concluded that the anomalies below T$_C$ is the signature of spin-phonon interaction arising from the paramagnetic to FM  transition. The larger spin-phonon interaction implies that atoms with stronger bonds have a greater influence on the magnetic ordering with colinear spin states. We attribute that below T$_C$ the structural fluctuation rearranges the spins and tends to frustrate spin states with larger magnetization. A schematic model, as shown in Figure \ref{model}(a) shows the normal lattice vibration and Figure \ref{model}(b) illustrates the lattice vibration with spin-spin interaction and distortion of the exchange coupling due to spin-phonon interaction due to laser shining. Both the Raman modes and the magnetic moment as a function of temperature strongly suggest the interaction of spin momentum with lattice phonons below T$_C$ due to nearest neighbor spin-spin interactions.

\section{Conclusions}

The quasi 2D FM $Cr_2Te_3$ flakes were synthesized and optimized from co-evaporated thin films and subsequent plasma treatment. The FESEM and EDX analysis confirms the surface flake-like structures. The XRD and HRTEM confirm the (001) oriented single crystal $Cr_2Te_3$ flakes with c-axis van-der-wall gap of 2.95 \AA. The magnetic behavior of the 2D $Cr_2Te_3$ has been investigated using SQUID measurements with both perpendicular and parallel applied magnetic fields to the c-axis of the flakes. A mixed magnetic phase with strong FM and weak AFM components has been confirmed from M-H  and M-T measurements. The polarization-dependent Raman analysis of the thin flakes provides insights into the modes of lattice vibration. Additionally, the DFT calculations support the experimental frequency of the Raman modes. This strengthens the identification of the Raman modes and supports the reliability of the experimental observations. The observed Raman frequency, FWHM, and integrated intensity from the temperature-dependent Raman measurements indicate the transition temperature. The anomalies below the T$_C$ suggest the strong spin-phonon interaction parameter ($\lambda_{sp}$) of 0.47$\pm$0.03 cm$^{-1}$. The combination of experimental techniques, along with theoretical calculations, has provided a comprehensive understanding of the structural, magnetic, and vibrational properties of the 2D $Cr_2Te_3$ flakes and strengthened the understanding of strong spin-phonon interaction mechanisms. The spin-phonon phenomena of 2D ferromagnetic $Cr_2Te_3$ thin flakes can be suitable for 2D magnetic spintronic device applications.

\begin{acknowledgments}

The authors acknowledge the National Institute of Science Education and Research Bhubaneswar, DAE, India for supporting this work through the project RIN-4001.

\end{acknowledgments}

\nocite{*}



\providecommand{\noopsort}[1]{}\providecommand{\singleletter}[1]{#1}%

\end{document}